\documentclass[12pt]{article}
\usepackage{epsf}

\newcommand{\bee}{\begin{equation}}
\newcommand{\ee}{\end{equation}}
\newcommand{\bea}{\begin{eqnarray}}
\newcommand{\eea}{\end{eqnarray}}
\newcommand{\R}{\rm I\kern-.2emR}
\newcommand{\C}{\rm \kern.25em\vrule height1.4ex
depth-.12ex width.06em\kern-.31em C}
\newcommand{\N}{{\rm I\kern-.16em N}}
\newcommand{\Z}{{\rm Z\kern-.35em Z}}

\begin{document}
\begin{flushright}
\end{flushright}
\bigskip\bigskip\begin{center}
{\Huge                                                                         
Comment on ``Scaling Hypothesis for the Spectral Densities in the
$O(3)$ Nonlinear Sigma Model''
}
\end{center}
\vskip 1.0truecm
\centerline{Adrian Patrascioiu}
\centerline{\it Physics Department, University of Arizona,}
\centerline{\it Tucson, AZ 85721, U.S.A.}
\vskip5mm
\centerline{and}
\centerline{Erhard Seiler}
\centerline{\it Max-Planck-Institut f\"ur Physik}   
\centerline{\it (Werner-Heisenberg-Institut)}
\centerline{\it F\"ohringer Ring 6, 80805 Munich, Germany}
\bigskip \nopagebreak 

\begin{abstract}
We comment on the recent paper by Balog and Niedermaier
\cite{bn}.
\end{abstract}
\vskip2mm

Balog and Niedermaier's scaling hypothesis for the spectral densities in
the $O(3)$ nonlinear $\sigma$ model \cite{bn} does not seem to be 
supported by the Monte Carlo data. As we showed in a recent paper 
\cite{dod}, the numerics indicate that the continuum limit of the lattice $O(3)$ 
$\sigma$ model agrees as well with the S-matrix prediction \cite{bn2} as
with the continuum limit of the dodecahedron spin model. In the latter  
model the massive high temperature phase must terminate at some finite
inverse temperature $\beta$. It is well known that asymptotic freedom
requires the current 2-point function to diverge for $p/m\to\infty$
\cite{bal}. However in another recent paper we proved that the current 
2-point function is bounded at finite $\beta$, due to reflection
positivity and a Ward identity.

\begin{figure}[htb]
\centerline{\epsfxsize=4.2cm\epsfbox{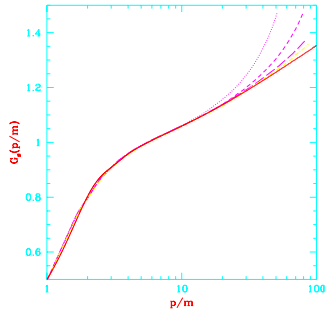}}
\caption{Monte-Carlo data and the Balog-Niedermaier prediction (solid 
line) for the spin two point function.}
\label{spino3log}
\end{figure}

\begin{figure}[htb]
\centerline{\epsfxsize=4.2cm\epsfbox{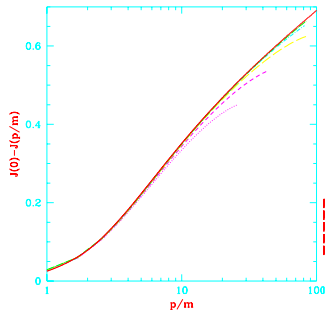}}
\caption{Monte-Carlo data and the Balog-Niedermaier prediction (solid
line) for the current two-point function.}
\label{curro3log}
\end{figure}

Consequently, unless the observed agreement between the $O(3)$ and 
dodecahedron spin model is accidental and disappears at larger
values of $p/m$, the Balog and Niedermaier scaling hypothesis, predicting
a logarithmic increase of the current 2-point function with $p/m$, 
cannot be correct. Moreover, since the spin 2-point function should
behave as $(p/m)^{\eta-2}$ for large $p/m$, the behavior
of the odd and even-particle number spectral densities $\rho^{(n)}(\mu)$
must be quite different.

This different behavior is shown both by our Monte Carlo data and
by the Balog-Niedermaier prediction itself, if one looks at it on a
logarithmic scale (see Fig.\ref{spino3log}, Fig.\ref{curro3log}):
while for the odd case (spin) the data start growing faster than
logarithmically, they  grow more slowly than $\ln(p/m)$ for the even case
(current). This behavior gives support to our scenario of a power-like
increase in the odd and boundedness in the even sector. Incidentally for 
the odd case the logarithmic slope at $p/m=100$ is already .143 and 
growing; this is larger than the prediction $4/3\pi^2$ in \cite{bn2} 
(the long version of their letter) for the asymptotic slope.

Finally, Fig.1 in \cite{bn2} is not representing our Monte Carlo data
for $p^2 G(p)$ (instead of $p^2$ a lattice variant is used); $p^2 G(p)$
is shown in Fig.\ref{spino3log} here (see also \cite{dod}).
Also the introduction of \cite{bn2} contains a misleading statement 
regarding our work: they give the incorrect impression that 
superinstantons are creating a mass gap. After pointing out, correctly, 
that superinstantons restore the $O(3)$ symmetry \cite{si},
they go on to state the triviality that in the absence of a KT transition,
the theory has a mass gap.  The way the sentences are juxtaposed,
a false logical connection is suggested.
In fact, via our percolation arguments
\cite{pat,perc}, the super-instanton gas is naturally associated with
masslessness and we believe that the massive, high temperature phase,
which appears to be correctly described by the Zamolodchikovs' 
S-matrix has nothing to do with the large $\beta$ regime of the model,
which is dominated by super-instantons and is massless.

\end{document}